\documentclass[10pt]{article}    
\usepackage[OE]{express}

\usepackage{amsmath,amssymb,graphicx}
\usepackage{caption}
\usepackage{graphicx}
\usepackage{color}
\usepackage[justification=raggedright]{caption}
\usepackage{ulem}
\usepackage{ragged2e}

\begin{document}

\title{Directional optical switching and transistor functionality using optical parametric oscillation in a spinor polariton fluid}

\author{Przemyslaw Lewandowski,\authormark{1} 
Samuel M. H. Luk,\authormark{2}
Chris K. P. Chan,\authormark{3}
P. T. Leung,\authormark{3}
N. H. Kwong,\authormark{4}
Rolf Binder,\authormark{2,4}
and
Stefan Schumacher \authormark{1,4,*}}

\address{\authormark{1} Physics Department and Center for Optoelectronics and Photonics Paderborn (CeOPP), Universit\"at Paderborn, Warburger Strasse 100, 33098 Paderborn, Germany  \\
\authormark{2} Department of Physics, University of Arizona, Tucson, AZ 85721, USA \\
\authormark{3} Department of Physics, The Chinese University of Hongkong, Hongkong SAR, China \\
\authormark{4} College of Optical Sciences, University of Arizona, Tucson, AZ 85721, USA}

\email{\authormark{*}stefan.schumacher@uni-paderborn.de} 








\begin{abstract}
Over the past decade, spontaneously emerging patterns in the density of polaritons in semiconductor microcavities were found to be a promising candidate for all-optical switching. But  recent approaches were mostly restricted to scalar fields, did not benefit from the polariton's unique spin-dependent properties, and utilized switching based on hexagon far-field patterns with 60$^\circ$ beam switching (i.e. in the far field the beam propagation direction is switched by 60$^\circ$). Since hexagon far-field patterns are  challenging, we present here an approach for a linearly polarized spinor field, that  allows for a transistor-like (e.g., crucial for cascadability) orthogonal beam switching, i.e. in the far field the beam is switched by 90$^\circ$. We show that switching specifications such as amplification and speed can be adjusted using only optical means.
\end{abstract}




\section{Introduction}
\label{sec:intro}
Exciton-polaritons in semiconductor microcavities are well-known for their unique optical properties.
Owing to a strong nonlinearity and a long coherence time,
they have been investigated intensely
not only in terms of their fundamental principles,
but also in regard to
possible optoelectronic applications \cite{schumacher-etal.09pssrrl,dawes-etal.10,Luk2013,Balarini2013,Niemitz2016, Amo2010, Steger2012, Marsault2015, Flayac2013, Sanvitto2016}. A promising system for all-optical switching  is that of stationary optical patterns 
%
%
in the polariton density \cite{schumacher-etal.09pssrrl,dawes-etal.10,Luk2013,Saito2013,Ardizzone2013,Egorov2014, Lederer2014b}, which can be viewed as solid-state version of the switching reported for gaseous media in
\cite{Dawes2005,Dawes2008}.
 Above a certain density threshold, a spatially homogeneous ensemble of coherently pumped polaritons can become optically unstable. Driven by a strong repulsive exciton-exciton interaction, pump polaritons start to scatter spontaneously into off-axis modes, breaking the system's translational symmetry and giving rise to an optical parametric oscillation (OPO). By their photonic part, these off-axis polaritons partially leak out of the cavity and are then detectable as additional emitted light beams in the far field.
Such  patterns and their optical control have been studied theoretically \cite{schumacher-etal.09pssrrl,dawes-etal.10,Luk2013, schumacher-etal.09pssrrl, Dawes2005, Dawes2008}, and
a  directional switching has been observed experimentally \cite{Ardizzone2013}, but amplification and transistor-like action has not yet been demonstrated. 

The motivation for the present work lies in the fact that 
the previous studies on this topic were  based on directional switching within a hexagon far-field pattern, where the switching results in a change of the beam direction
by 60$^\circ$. Since hexagonal pattern formation is more challenging than that of two-spot patterns, we present in the following a scheme that does not rely on (and indeed not use) any hexagonal pattern. Moreover, the
previous switching studies used mainly
 scalar fields and did not benefit from  spin-dependent properties, which lead to (i) the optical spin-Hall effect, where the polariton's polarization state is affected already in the linear optical regime \cite{Kavokin2005b}, (ii) a polarization dependent four-wave mixing resonance in the nonlinear regime below the instability threshold \cite{OmblinePaper}, and (iii) - of particular interest here - spin-dependent signatures on the optical patterns forming above the instability threshold \cite{Ardizzone2013,Egorov2014}. These phenomena are based mainly on two underlying mechanisms: (i) Stemming from the photonic part, polaritons have slightly different dispersions for linear polarization states oriented transverse and longitudinal to the light's plane of incidence (TE-TM splitting); (ii) a repulsive  (possibly attractive) interaction between excitons with parallel (opposite) spins governs a spin-dependent polariton-interaction in the nonlinear regime.
In particular, linearly polarized polaritons get strongly affected by both mechanisms: TE-TM splitting lifts the azimuthal symmetry, and approaching the instability threshold, pump polaritons scatter preferably into modes oriented parallel or perpendicular to the pump's polarization plane \cite{OmblinePaper}. This provides a natural platform for orthogonal-direction (i.e., 90$^\circ$) switching, as will be shown below.
An important advantage of the 90$^\circ$ switch over the previously studied 60$^\circ$ switch is that the former is not hindered by the TE-TM splitting but actually makes active use of it. In contrast to the 60$^\circ$ switch,  both incoming light fields (pump and control) are at the same frequency and these are the optimum conditions for the switching to occur.

%
\begin{figure}
\centering{\includegraphics[height=7cm]{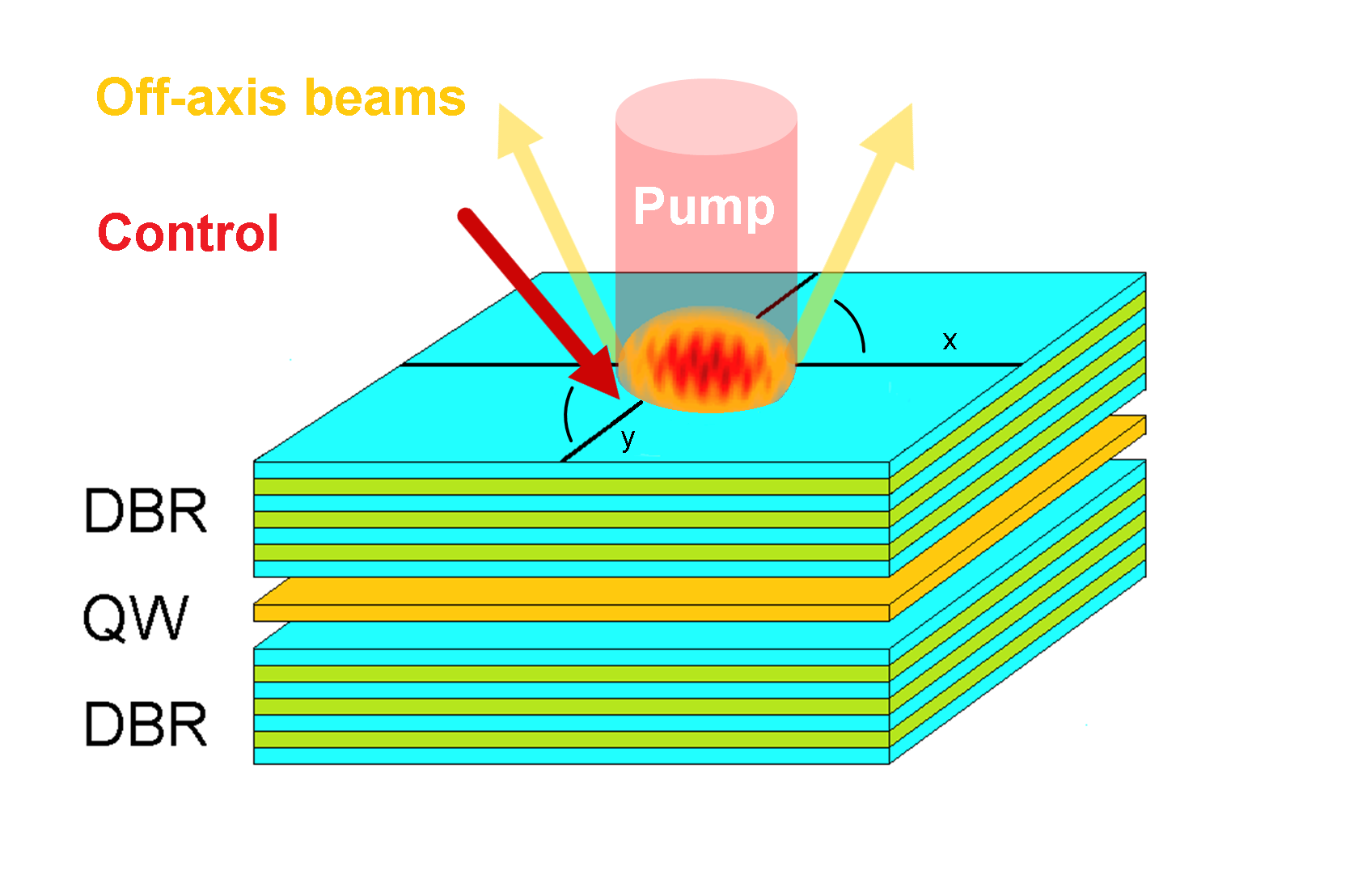}}
\caption{Sketch of the system and excitation geometry: a quantum well (QW) is embedded between two DBR mirrors, forming a semiconductor microcavity.  A linearly polarized pump beam is sent onto the cavity in normal incidence and excites a coherent polariton field. Above the instability threshold, the cavity spontaneously emits two off-axis beams with opposite in-plane momentum parallel to the pump's polarization plane. This initial far-field pattern can be switched by a weak control beam, polarized orthogonal to the pump polarization, entering the cavity in oblique incidence in a direction 90$^{\circ}$ rotated with respect to the initial emission direction.
}
\end{figure}

Figure 1 shows a sketch of the system and the excitation geometry: a linearly polarized cw-pump in normal incidence excites an ensemble of coherent polaritons. If the polariton density exceeds the instability threshold, the above-mentioned occupation of off-axis modes appears. For pump powers only slightly above the threshold, the pattern formation results in two outgoing beams. Both are cross-linearly polarized to the pump and leave the cavity in a direction parallel to the pump's linear polarization. 
Increasing the pump power further, this two-spot pattern could possibly transform into a hexagonal pattern as studied in \cite{Ardizzone2013,Egorov2014}, but such hexagon patterns have turned out to be challenging, especially in experiments.
We note that the  ``parallel'' state of the initial 2-spot pattern, i.e.\ the emission direction being in a direction parallel to the pump's linear polarization,
is preferred over the ``perpendicular'' state, i.e.\ the emission direction being in a direction  perpendicular to the pump's linear polarization,
 because of a slightly different density of states for TM- and TE-modes.

A weak, cross-linearly polarized external beam is sent into the cavity with the same  angle of incidence as the emission of the outgoing beams, but with a plane of incidence rotated by $90^\circ$ relative to the plane of the outgoing beams. 
 This ``control beam'' stimulates a pairwise, phase-matched scattering of pump polaritons into off-axis modes parallel to its incidence plane and ``switches on'' an optical two-spot pattern in this direction. For sufficiently strong control beams, the initial pattern becomes unstable and is switched off. A deeper understanding of the non-equilibrium phase transitions on the basis of a population-competition model has been given in \cite{tse-etal.15} for the hexagonal geometry. 
Switching off the control beam, the OPO in the former direction as preferred by the system's anisotropy emerges again and the initial pattern returns - the switching is reversible. The strength of the anisotropy can easily be enhanced by tilting the pump slightly along the preferred direction: this allows for tailoring of switching attributes like possible gain or time scale using only optical means.
\begin{figure}
\centering{\includegraphics[height=7cm]{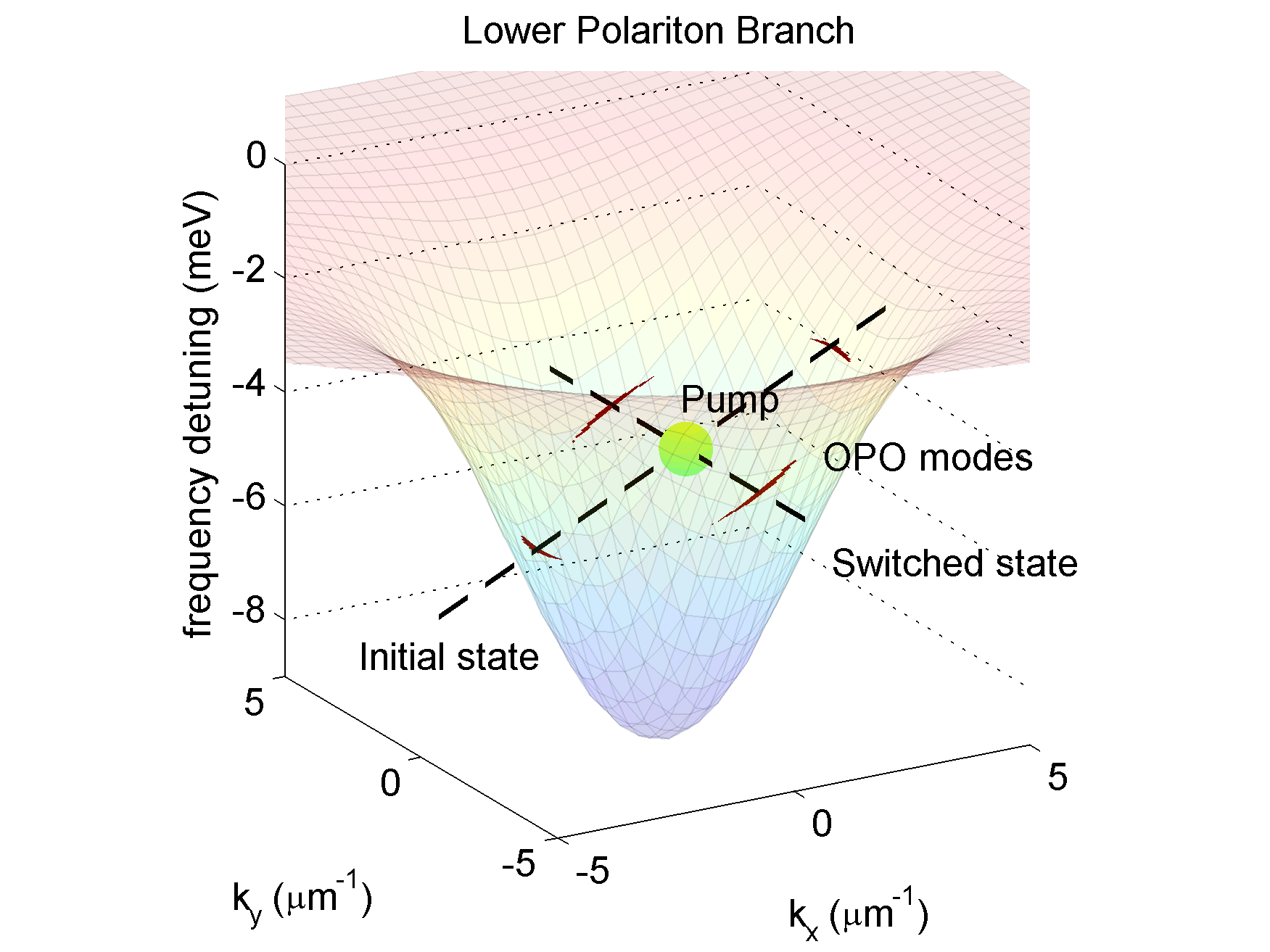}}
\caption{Lower polariton branch in the two-dimensional momentum space and excitation scheme: pump-polaritons with finite momentum are excited 4 meV above the polariton ground-state. Modes on the LPB eligible for a self-sustained OPO are marked red. The spontaneous emergence  of  the preferred two-spot pattern oriented along the pump's tilting axis (x-axis) can be switched off triggering an OPO in the y-direction.}
\end{figure}

Figure 2 shows the dispersion of the lower polariton branch (LPB) in the two-dimensional momentum space ($k_x$, $k_y$). For the sake of clarity, the splitting in TE- and TM-branch is not shown here. The $x$-polarized pump excites the polaritons off-resonantly 4 meV above the polariton ground state and with a finite momentum $\bold{k}_\mathrm{pump} = (k_\mathrm{pump},0)$ in $x$-direction. The polariton density is adjusted to be slightly above the instability threshold. Using a linear stability analysis \cite{OmblinePaper, Schumacher2007a, Schumacher2007c}, we are able to determine an ``effective decay rate'' $\gamma_\mathrm{eff} (\bold{k})$ for each mode, considering not only the intrinsic loss rate $\gamma$, but also the amplification through stimulated scattering of  pump polaritons into the given mode. If this amplification exceeds the intrinsic losses, any initial polariton field in this mode can start growing exponentially, giving rise to a spontaneous pattern formation as mentioned above. In Fig. 2, the modes eligible for such a self-sustained OPO are marked red: the pump-polaritons can scatter spontaneously either pairwise in $x$- or in $y$-direction (parallel or perpendicular to the pump's polarization plane, respectively). The x-direction  is favored through (i) a slightly higher density of state for TM-modes and (ii) a finite tilt of the pump-beam. To fulfill phase-matching and resonance conditions simultaneously, the polaritons undergo slight frequency shifts in the course of scattering in this direction. The control beam shown in Fig. 1 excites modes eligible for an OPO oriented along the y-axis and, importantly, is at the same frequency as the pump beam.

\begin{figure}
\centering{\includegraphics[scale=0.18]{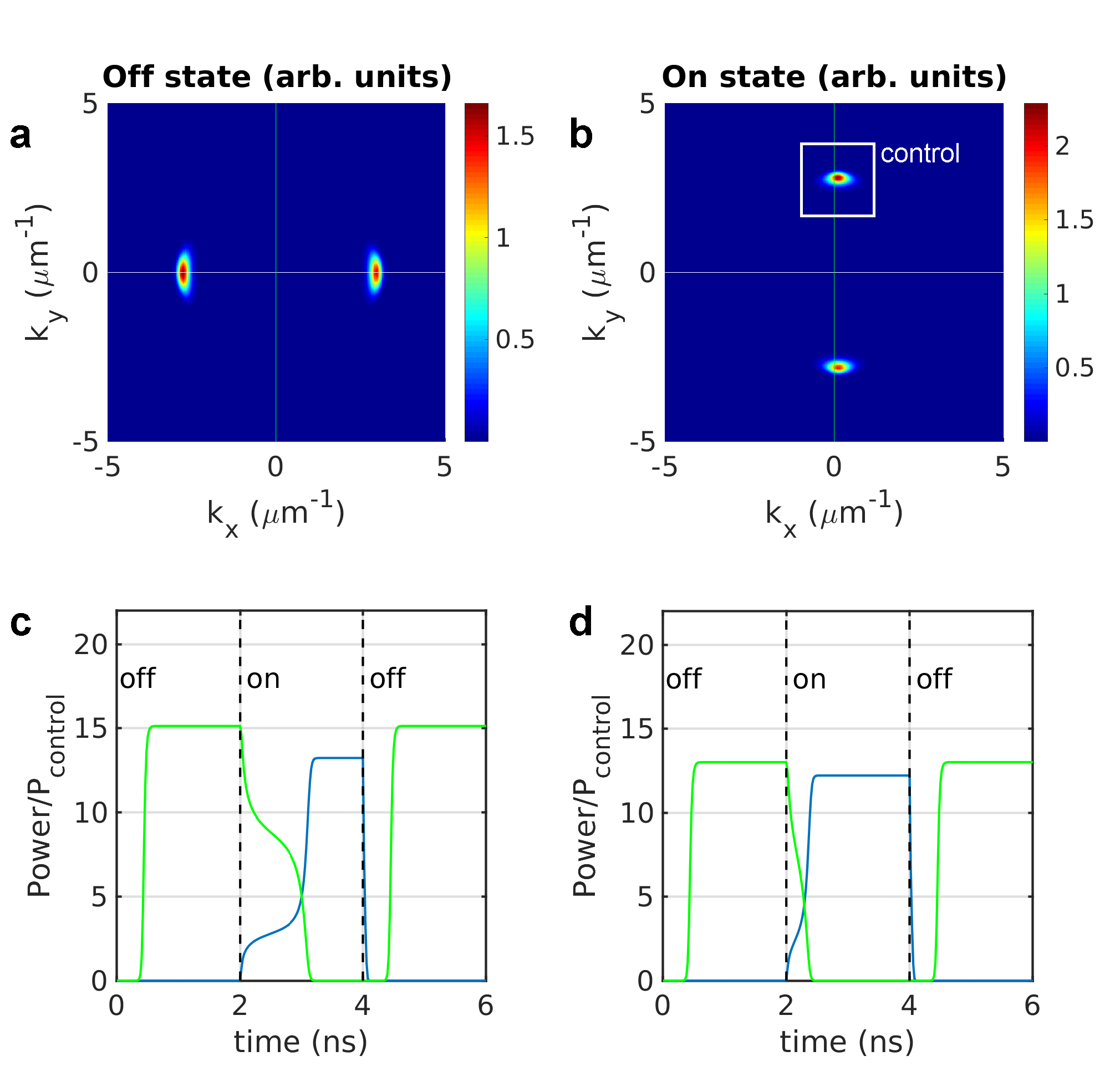}}
\caption{
The initial far-field pattern, emerging in the cross-linear polarization channel, is shown in the momentum space (a). Switching on the control beam (indicated by a white square), a two spot-pattern emerges in the y-direction, and the initial pattern is suppressed (b). The intensity of the initial state (green) and of the target state (blue) during the switching is shown for a low and a high control intensity in c and d, respectively. The switch on and off times of the control beam are marked by vertical lines.}
\end{figure}

\section{Results}
The pattern formation and all-optical switching discussed above can be studied numerically, computing the polariton's dynamic in the real-space and time domain. Details
of the theoretical approach and the system's parameters are given below in the methods section. For a pump intensity slightly above the instability threshold, Fig. 3a shows the density of a stationary far-field pattern in the momentum space {\color{black} 2 ns} after the onset of pattern formation. The pump-polaritons are excited 4 meV above the polariton ground-state and with a finite momentum $k_\mathrm{pump} = 0.1$  $\mu$m$^{-1}$ as depicted in Fig. 2. The far-field pattern consisting of two spots is oriented parallel to the pump's tilt direction and hence to the pump's polarization plane, and emerges
 resonantly on the LPB in the cross-linear polarization channel.

This initial state can be switched by a control beam sent onto the cavity in oblique incidence, with the same frequency as the pump, but resonant on the LPB. The plane of incidence of this additional incoming beam is perpendicular to the orientation of the initial pattern, as shown in Fig. 3b. The control beam stimulates a pairwise scattering of pump polaritons. One pump polariton scatters into the direction of the control beam and amplifies it. To fulfill the phase-matching condition, a second polariton scatters in the opposite direction and gives rise to a field with opposite in-plane momentum. After a certain time - the ``switching time'' - this process results in the ``target state'' of the switching - a two-spot pattern aligned perpendicular to the initial one. With an OPO triggered externally, the  OPO in the initial direction is suppressed and the initial pattern switched off.
The applied control power amounts to {\color{black} $P_\mathrm{control} = 7.1$ $\mu$W} and is low compared to the intensity of the initially outgoing beams {\color{black} (each 99.7 $\mu$W)}. That implies a transistor-like switching with a high gain factor of about {\color{black}14}.

To shed light on the switching dynamics, Fig.\ 3c shows the power (in multiples of $P_\mathrm{control}$) of the pattern spots in the initial and target state in the course of the switching process. The initial pattern converges after approximately {\color{black}600 ps} after the onset of pattern formation. At {\color{black}$t=2$ ns}, the control beam is applied and triggers the all-optical switching, which is completed at $t=3.3$ ns after 1.3 ns. At {\color{black}$t=4$ ns}, the control beam is switched off, the target state collapses immediately and the initial pattern returns spontaneously after 600 ps.
Compared to the short back-switching time of {\color{black}0.6 ns}, an on-switching time of {\color{black} 1.3 ns} appears rather long. A possibility to accelerate the switching is to increase the control beam intensity, as done for Fig. 3d: a brighter control beam leads to a faster switch on a timescale below {\color{black}500 ps} and moreover to an intensity increase of the target pattern. %
This agrees with the intuition gained from a switching analysis based on non-equilibrium phase transitions, Ref.\  \cite{tse-etal.15}.
However, with a higher control beam power {\color{black}$P_\mathrm{control} = 8.2 $ $\mu$W} the gain factor is reduced {\color{black}(down to about 12)}, but the switching still remains distinctly transistor-like.

\begin{figure*}
\centering{\includegraphics[scale=0.20]{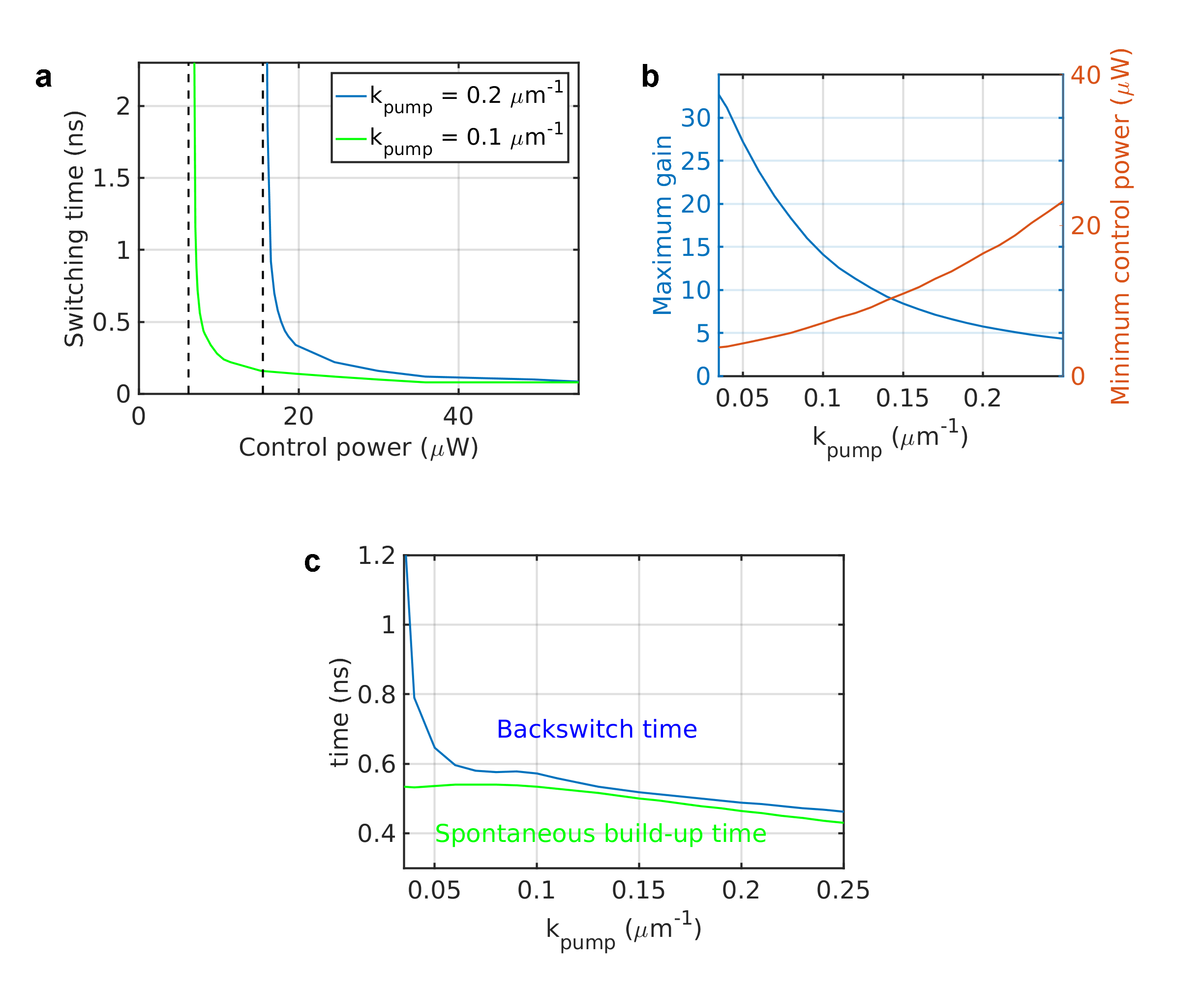}}
\caption{(a) Dependence of  switching time  on the applied control intensity for systems with two different pump momenta. Each minimum control intensity is marked by a vertical line. 
(b) Dependence of maximum gain and minimum control intensity on pump momentum (tilt away from normal incidence). (c) 
Back switch time after switching off the control beam (blue line) and spontaneous built-up time of the initial pattern (green line) as a function of pump tilt.}
\end{figure*}

As is apparent from Figs.\ 3c and d, the switching speed depends strongly on the applied control beam. To discuss this dependency in more detail, we show in Fig.\ 4a  the dependence of the switching time on the control power $P_\mathrm{control}$ for the current system (green line). As visible there, the switching time exhibits a threshold behavior: below a certain  minimum power {\color{black}($P_\mathrm{control,min} = 6.2$ $\mu$W )}, no complete switching is possible, since the applied beam is too low in power to make the initial pattern unstable and switch it off. Approaching this threshold from higher powers, however, the  switching time diverges. With larger intensities it decreases as expected and converges to a minimum at {\color{black}$t=80$ ps}. The blue graph in Fig.\ 4a shows a similar time-power dependence for a system driven with a higher pump momentum ($k_\mathrm{pump} = 0.2$ $\mu$m$^{-1}$). With a pump tilted farther away from normal incidence, the system's anisotropy is increased, and the minimum control power shifts to a higher value {\color{black}($P_\mathrm{control,min} = 15.5$ $\mu$W)}. The minimum switching time, however, does not change significantly.

The dependence of the minimum control beam power on $k_\mathrm{pump}$ and hence on the system's anisotropy, is shown in Fig.\ 4 b. The power required to switch the system's initial pattern increases approximately quadratically with the pump momentum. The power of each initial pattern, however, is almost constant for each $k_\mathrm{pump}$. As a consequence, the gain factor is inverse to $P_\mathrm{control,min}$ and decreases dramatically with increasing anisotropy (but is still greater than four in the momentum range studied here).

In contrast to the switching time (the time required to suppress the initial pattern), the back switch time (i.e., the time it takes for  the initial pattern to return after switching off the control)
is approximately independent of the control beam intensity, but strongly dependent on the pump momentum, as depicted in Fig. 4c (blue line). For pump momenta below {\color{black} $k_\mathrm{pump} = 0.035$ $\mu$m$^{-1}$} the system remains in the switched state - a back switch into the initial state does not occur. In the limit of high pump momenta, the back switch time decreases - the stronger the anisotropy, the faster the system recovers the preferred initial state. In this regime, the back switch time is slightly larger than the time necessary for the spontaneous build-up of the initial pattern (green line). Approaching {\color{black} $k_\mathrm{pump} = 0.035$ $\mu$m$^{-1}$} from higher momenta, the back switch time diverges since the initial state (two-spot pattern) is not sufficiently stable.
The transient dynamics show a nontrivial behavior with a small local maximum of the backswitch time at {\color{black} $k_\mathrm{pump} = 0.08$ $\mu$m$^{-1}$} .
\section{Conclusions}
\label{sec:conclusion}
We have presented the concept of an all-optical switch in which the direction of the far-field emission from a semiconductor microcavity is rotated by 90$^\circ$. The physics of the switching mechanism involves optical parametric oscillation and non-equilibrium phase transitions of polariton patterns. Furthermore, the 90$^\circ$ switch utilizes the spin-dependent polariton interactions and the cavity's TE-TM splitting. Our theoretical analysis shows that the on and off switching times have different parametric dependencies.  The switch-on time can be reduced by increasing the control beam intensity, and the switch-off time can be reduced by increasing the pump tilt. Nanosecond or sub-ns switching times are possible. The gain, defined as the switched power over the control power, is found to be between one and two orders of magnitude. The specifications of the switching can be adjusted using optical means. Further analysis and optimization, for example using novel cavity design concepts  \cite{kim-etal.2016}, and experimental realization are planned for the future.

\section{Appendix -- Methods}
We compute the coherently pumped polariton field solving the coupled dynamics of the photonic and excitonic constituents. For a single cavity system, our approach is based on a quasi-mode approximation for the optical field $E^\pm$ and a microscopic semiconductor theory for the excitonic polarization field $p^\pm$ \cite{OmblinePaper}, where $\pm$ denotes the circular polarization states. The dynamics of those fields is given by
\begin{eqnarray}
i \hbar \frac{\partial}{\partial t} E^\pm  & = & \left( \mathbb{H} - i \gamma_c \right) E^\pm +  \mathbb{H}^\pm E^\mp - \Omega_x p^\pm + E^{\pm}_\mathrm{pump} \nonumber \\
i \hbar \frac{\partial}{\partial t} p ^\pm & = & \left( \epsilon^x_0 - i \gamma_x \right) p^\pm - \Omega_x \left(1 - \alpha_\mathrm{PSF} \vert p^\pm \vert^2\right) E^\pm \nonumber \\
& & + T^{++} \vert p^\pm \vert^2 p^\pm + T^{+-} \vert p^\mp \vert^2 p^\pm.\nonumber
\end{eqnarray}
Here, $\mathbb{H}=\epsilon^c_0 -\frac{\hbar^2}{4}\left( \frac{1}{m_\mathrm{TM}}+ \frac{1}{m_\mathrm{TE}}\right) \left( \frac{\partial^2}{\partial x^2} + \frac{\partial^2}{\partial y^2}\right)$ denotes the free-particle Hamiltonian, $\mathbb{H}^\pm = -\frac{\hbar^2}{4}\left( \frac{1}{m_\mathrm{TM}}- \frac{1}{m_\mathrm{TE}}\right) \left( \frac{\partial^2}{\partial x^2} \mp i \frac{\partial}{\partial y}\right)^2$ is an off-diagonal element stemming from the TE-TM splitting, and $\Omega_x = 6.5$ meV denotes the photon-exciton coupling. $E^{\pm}_\mathrm{pump}$ is the external pump source. The loss rates of photons and excitons are $\gamma_c = 0.4$ meV and $\gamma_x = 0.4$ meV, respectively. The effective masses for TM and TE modes are $m_\mathrm{TM} = 0.215$ meV ps$^2$ $\mu$m$^{-2}$ and $m_\mathrm{TE} = 1.05 \cdot m_\mathrm{TM}$, respectively. The excitonic energy dispersion, here assumed as flat dispersion, is given by $\epsilon^x_0$. The photonic ground state is detuned from the exciton by $\epsilon^c_0 = \epsilon^x_0 - 5$ meV. The nonlinear interaction constants are given by a repulsive interaction $T^{++} = 5.69\cdot 10^{-3}$ meV $\mu$m$^{2}$ for excitons with parallel spin, and an attractive interaction $T^{+-} = -T^{++}/3$ for excitons with opposite spin. The phase-space filling constant is $\alpha_\mathrm{PSF} = 5.188 \cdot 10^{-4}$ $\mu$m$^{2}$.

After solving the coupled equations in real space and time domain, the results are transformed in the momentum space. The pump intensity outside of the cavity is related to $E^{\pm}_\mathrm{pump}$ by $I^\pm_\mathrm{pump} = \omega_\mathrm{pump} \gamma_c^{-1} \vert E^\pm_\mathrm{pump} \vert^2$, and the intensity of the outgoing photonic field inside the cavity is $I^{\pm} = \omega_\mathrm{pump} \gamma_c \vert E^\pm \vert^2$, where $ \hbar \omega_\mathrm{pump} = 1.5$ eV denotes the pump frequency.  To obtain the power of the outgoing pattern and control beams ($P_\mathrm{out}$ and $I_\mathrm{control}$, respectively), the intensities are integrated over real space. The gain factor $g$ can be obtained dividing the output pattern power by the power of the external control beam, $g = P_\mathrm{out}/P_\mathrm{control}$.

%
%
%
%
%
%
%
%
%
%

%

\section*{Funding}
The Paderborn group acknowledges financial support from the DFG (SCHU 1980/5 and TRR 142) and a grant for computing time at $\mathrm{PC^2}$ Paderborn Center for Parallel Computing. Stefan Schumacher further acknowledges support through the Heisenberg program of the DFG.
The Arizona group acknowledges financial support from NSF under grant ECCS-1406673 and from TRIF SEOS.
%
%

%





\end{document}